\documentclass[letterpaper, 10 pt, conference]{cssconf}  

\IEEEoverridecommandlockouts                              

\usepackage{epsfig} 
\usepackage{times} 
\usepackage{amsmath} 
\usepackage{amssymb}  
\usepackage{caption}
\usepackage{subcaption}
\usepackage{wrapfig}
\usepackage{multicol}
\usepackage{url}

\title{\LARGE \bf  Human Robot Interface for Assistive Grasping}

\author{David Watkins-Valls$^1$, Chaiwen Chou$^1$, Caroline Weinberg$^1$,  Jacob Varley$^1$, Kenneth Lyons$^3$, Sanjay Joshi$^3$ \\ Lynne Weber$^2$, Joel Stein$^2$, Peter Allen$^1$
	\thanks{This work was supported by NSF Grant 1208153. The authors are from $^1$Department of Computer Science, Columbia University $^2$Department of Rehabilitation Medicine, Columbia University and $^3$Department of Mechanical and Aerospace Engineering, UC Davis.}
}

\begin{document}

\maketitle
\thispagestyle{empty}
\pagestyle{empty}

\begin{abstract}
	This work describes a new human-in-the-loop (HitL) assistive grasping system for individuals with varying levels of physical capabilities. We investigated the feasibility of using four potential input devices with our assistive grasping system interface, using able-bodied individuals to define a set of quantitative metrics that could be used to assess an  assistive grasping system. We then took these measurements and created a generalized benchmark for evaluating the effectiveness of any arbitrary input device into a HitL grasping system. The four input devices were a mouse, a speech recognition device, an assistive switch, and a novel sEMG device developed by our group that was connected either to the forearm or behind the ear of the subject. These preliminary results provide insight into how different interface devices perform for generalized assistive grasping tasks and also highlight the potential of sEMG based control for severely disabled individuals. 
\end{abstract}

\section{Introduction}
This paper describes contributions towards the implementation of a Human-in-the-Loop (HitL) grasping system for assistive robotics. Although progress in the field of robotics has been swift, it is unlikely that truly independent operation of robots in situations where they will interact closely with objects, obstacles, and people in their environment will evolve in the immediate future. However, with the help of a human operator, it is possible to achieve robust, safe operation in complex environments. This work describes a system which can accomplish this goal with minimal interfaces that are accessible even to individuals with physical impediments, which will enable the development of more capable assistive devices for these individuals. The interface was designed to be compatible with a variety of input devices in order to give a more robust set of interfaces to a human user. 

Grasping an object generally requires contextual knowledge of the object and the intent of the user, particularly in cluttered scenes. We have developed a user interface that allows the user to effectively express their intent. This interface is validated by testing users with several input devices - a surface electromyography (sEMG) device, a mouse, an assistive switch, and an Amazon Echo Dot. The sEMG device and the assistive switch can be calibrated very easily, and both come with an enhanced GUI that improves user experience and the capability of the system. This work forms the foundation for a flexible, fully featured HitL system that will allow users to grasp objects using a variety of interface devices that not only has the potential to bring HiTL assistive devices out of the research environment and into the lives of those that need them, but also forms a metric for evaluating the effectiveness of both the task given to human subjects as well as the ease-of-use of the inputs into the system. 

\section{Related Work}
\begin{figure}[t]
	\centerline{
		\includegraphics[width=0.45\textwidth]{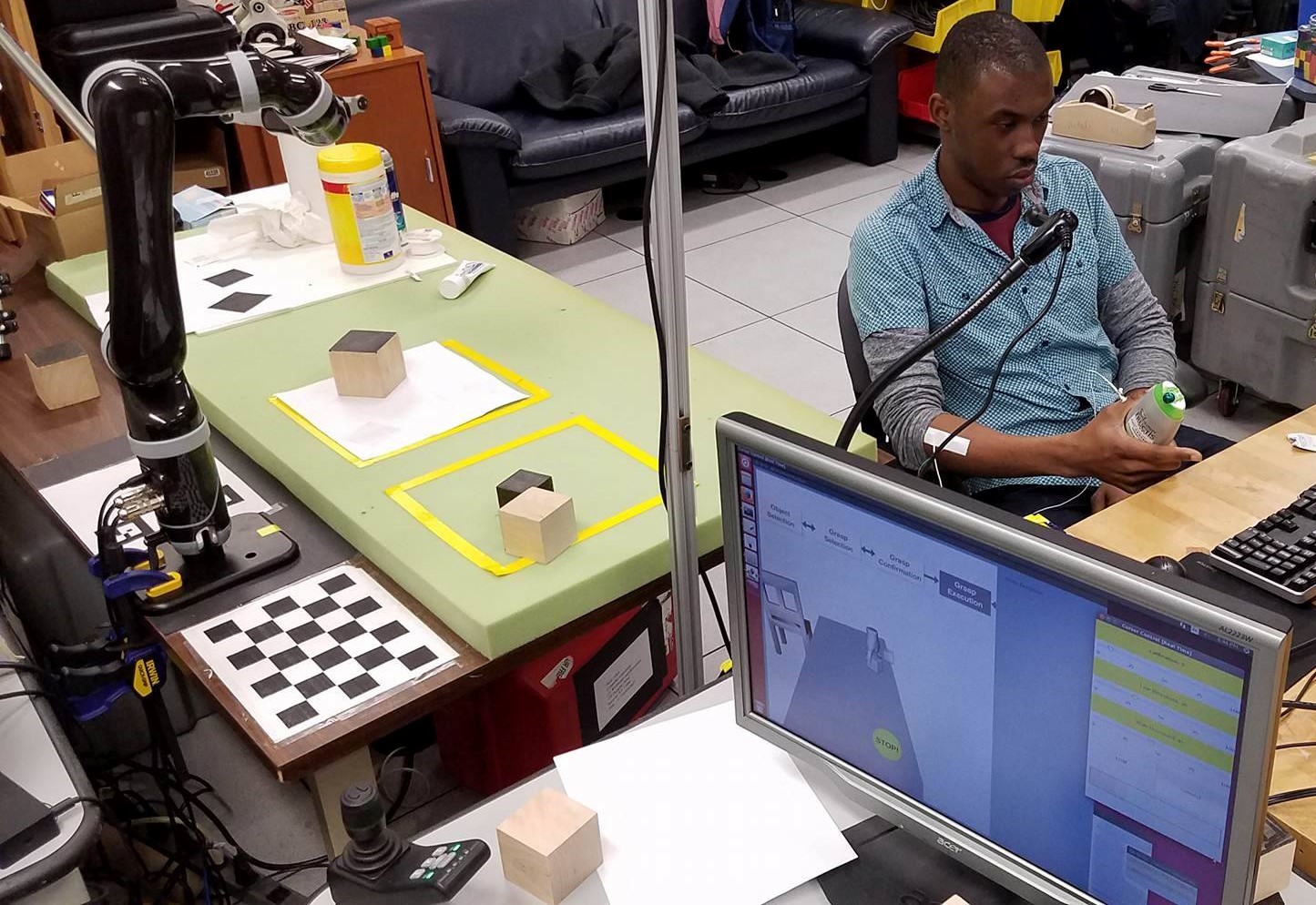}
	}
	\caption{Experimental setup shown during a grasp execution while using the sEMG device attached to the subject's forearm. The user interface, Kinova MICO robotic arm, Kinect depth sensor, sEMG device, and target objects are all visible.}
	\label{setup}
\end{figure}

As assistive robotics has developed, various types of control interfaces have been explored. The modalities of these interfaces range from devices that provide a binary signal (such as a sip-and-puff) to speech recognition to a brain-computer interface (BCI) \cite{Krishnaswamy2016}. Recent work in designing a HitL system that specializes in path planning under interface based constraints utilizes a sip-and-puff to allow for multiple inputs based on the state of the current interface. This was shown to be effective in assisting those with lower motor capabilities move around a room without requiring elaborate control over their wheelchair\cite{ICAPS1613145}. 

Other work in speech recognition in the context of assistive robotics has shown that many of the challenges facing assistive systems focus on making sure the system can understand the user and create an engaging system that performs the designated task well\cite{Rudzicz:2015:SIP:2785580.2744206}.

Not only is the form of input into the system important, but so is the design of the interface. Much work and care must be put into making sure the system is robust to user input and also allows the user to accurately see what is happening in the environment as a result of the system\cite{6385907}. Because HitL software is so focused on getting accurate input from the user, there is not much room for ambiguity\cite{7518989}.

In order to evaluate the effectiveness of this system we used a pick and place task that was developed as a modification of the Box and Blocks Test of manual dexterity\cite{doi:10.5014/ajot.39.6.386}. This test is used to evaluate physically handicapped individuals to develop a normative metric for adults. We believe a test such as this is appropriate for evaluating whether the system developed is achieving a useful and/or meaningful task. We opted for this instead of the Action Research Arm Test (ARAT) because it requires less than 1cm of noise for detecting the objects to be manipulated in that test and the equipment required for such precise point cloud measurements is not cost effective for an ordinary user at the time of this writing. 

\section{An HRI Grasping Platform}
\begin{figure}[t]
	\vspace{0.1in}
	\includegraphics[width=0.24\textwidth]{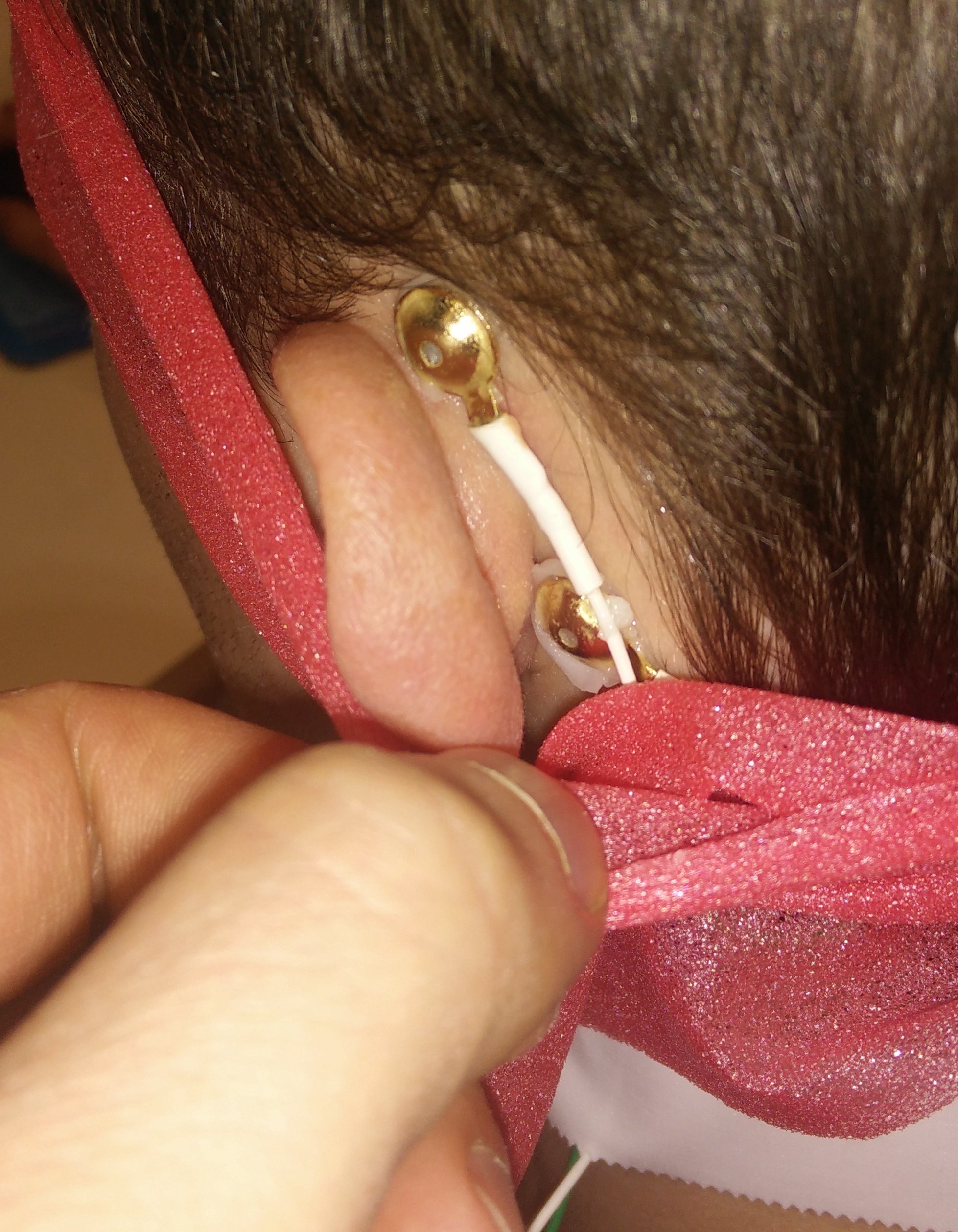}
	\includegraphics[width=0.24\textwidth]{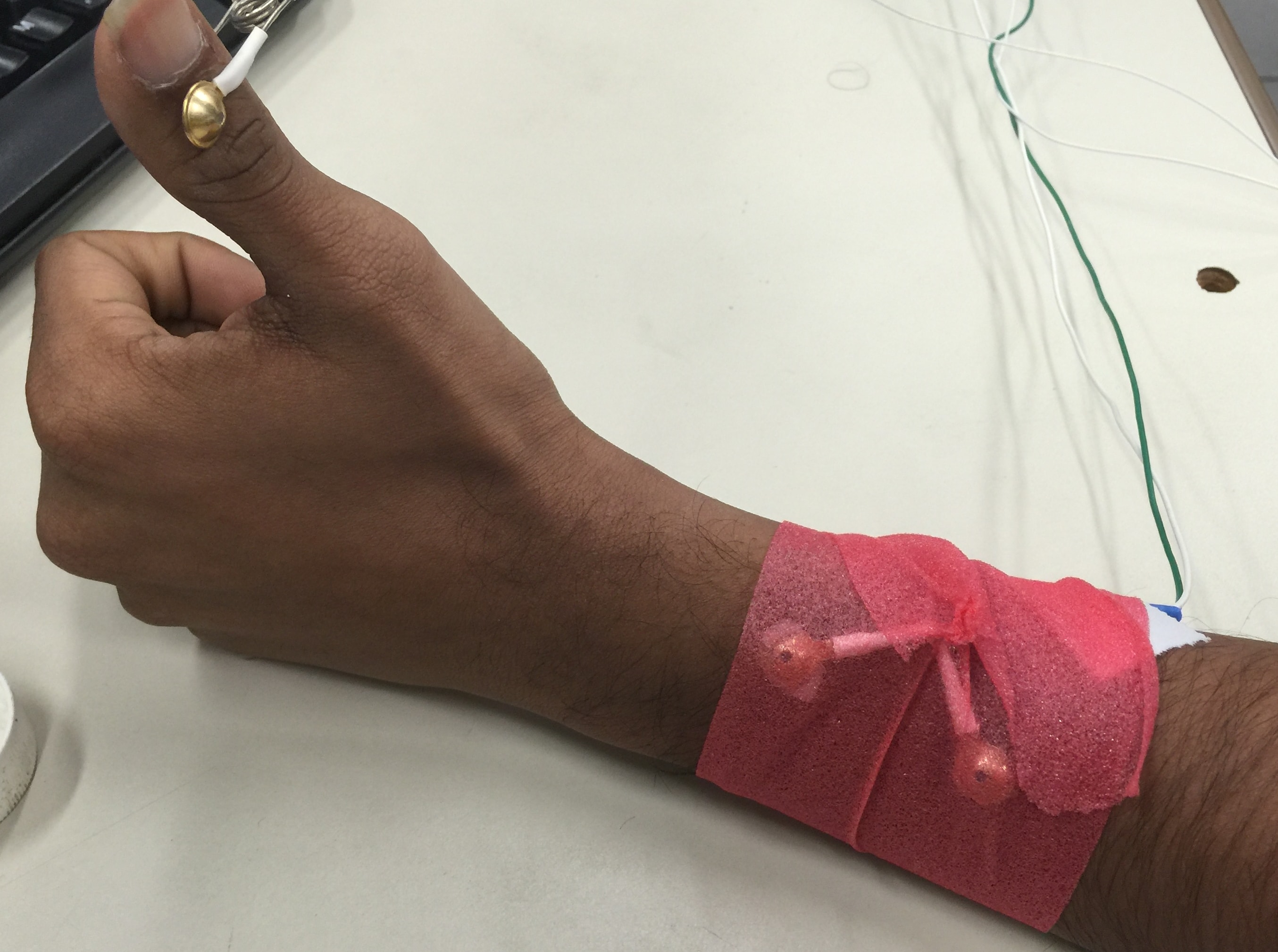}
	\caption{For the user studies, the sEMG device was placed over the superior auricularis (SA) muscle, located behind the ear for 3 subjects, and over the extensor pollicis longus (EPL) muscles, located on the forearm for the other 3 subjects.  The electrodes were attached with Ten20 conductive gel, and a small amount of prewrap (red) was used to stabilize them. }
	\label{fig:semg_device_worn} 
\end{figure}
\begin{figure}[t]
	\vspace{0.1in}
	\centering
	\includegraphics[width=0.35\textwidth]{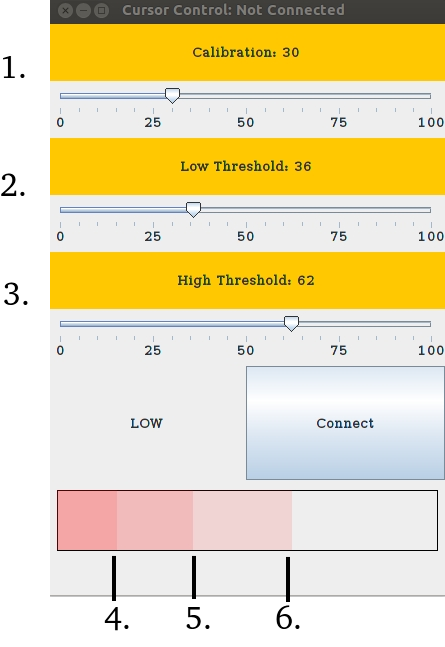}
	\caption{The sEMG interface contains a calibration slider (1) to set the gain applied to the sEMG signal. The low threshold slider (2) sets the threshold for the power bar that the user must exceed to change the currently selected button in the UI. The high threshold slider (3) sets the threshold for the power bar that the user must exceed in order to select a button in the UI. The power bar displays the current power output by the device (4) compared to the current low (5) and high (6) thresholds. } 
	\label{fig:device_interface}
	
\end{figure}
\begin{figure}[!ht]
	\centering
	\includegraphics[width=0.45\textwidth]{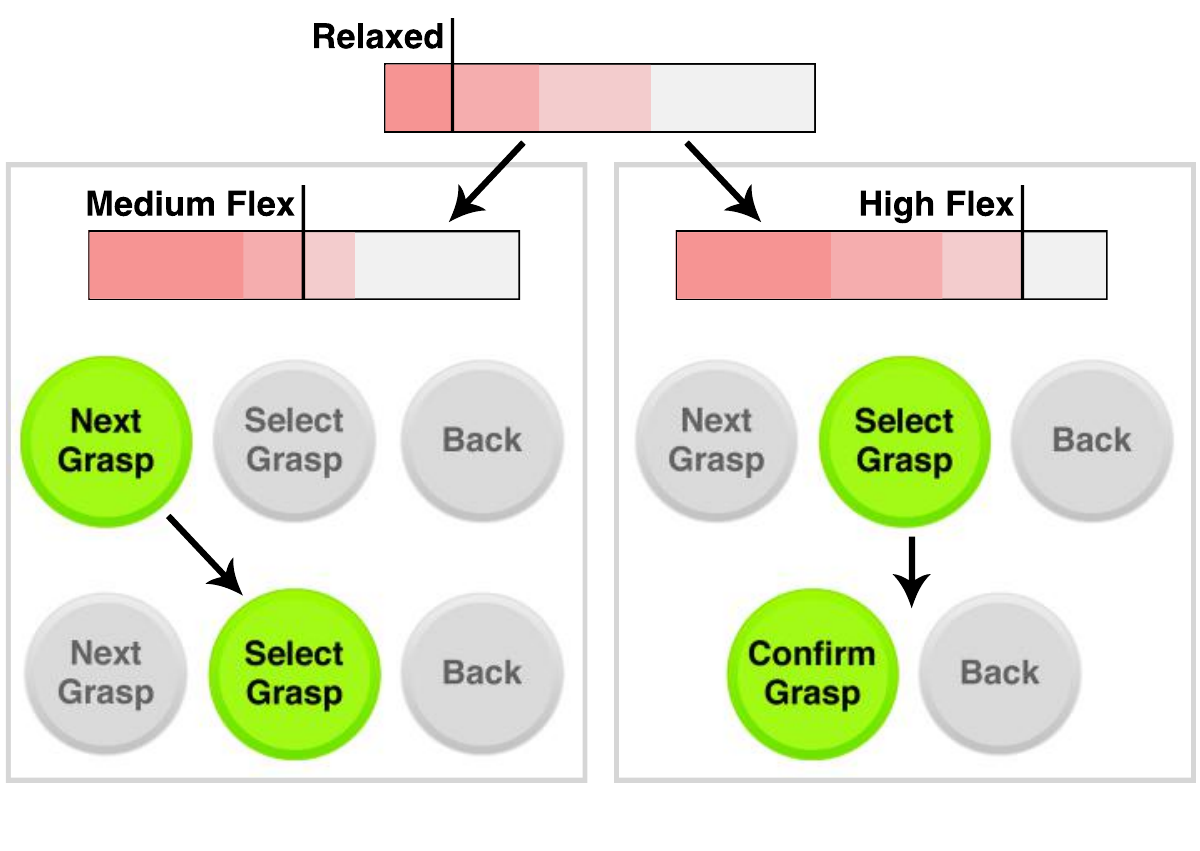}
	\caption{While the subject is in a relaxed state, no changes occur in the UI.  The subject is able to change the currently highlighted button with a medium flex, or select the currently highlighted button with a high or strong flex.}
	\label{fig:BCI_selection} 
\end{figure}

\begin{figure}
	\centering
	\includegraphics[width=0.41\textwidth]{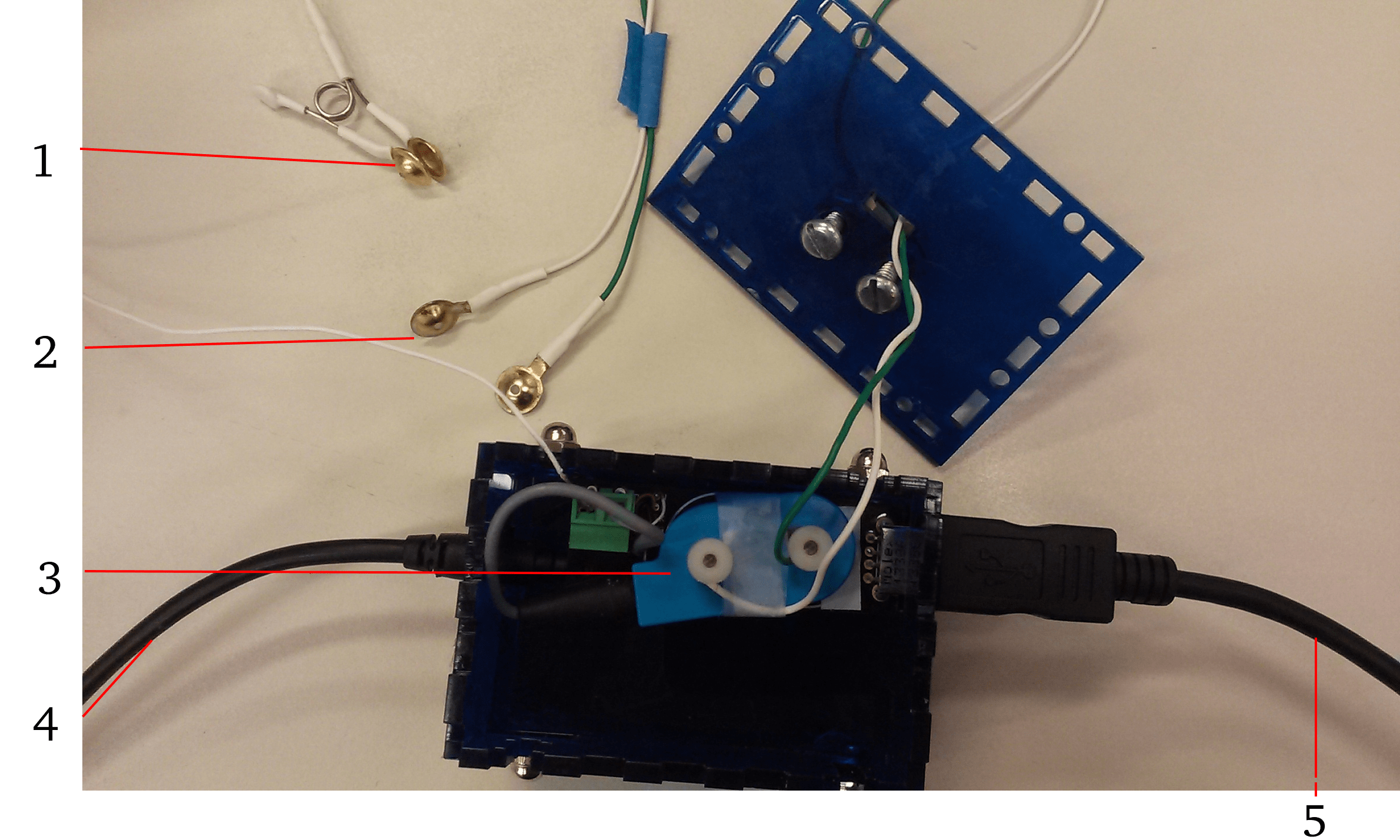}
	\caption{The sEMG device consists of a ground wire (1), 2 electrodes (2), amplifier (3), audio cable to connected to a microphone jack (4), and a USB cable to provide power (5). }
	\label{fig:device_semg}
\end{figure}

\begin{figure}
	\centering
	\includegraphics[width=0.41\textwidth]{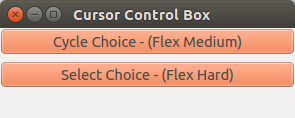}
	\caption{The mouse interface features two buttons. `Cycle Choice' cycles the highlighted option on the screen. `Select Choice' executes the highlighted command.}\label{fig:device_mouse}
\end{figure}

Our system is based on previous work that explored a number of different sEMG sensing devices and user interfaces (UI) which allow HiTL grasping \cite{Weisz2013}\cite{Weisz2013a}\cite{Weisz2014}.  Our work in developing those systems has outlined areas where improvements are needed, and below we describe each of these areas and the enhancements made to address them. Fig. \ref{setup} shows the experimental setup.  We are using a Kinova lightweight MICO arm, which has a two-fingered gripper, and a Microsoft Kinect RGB-D camera to provide point clouds of the scene. To speed up grasp planning in our constrained environment, we use a small set of grasps designed specifically for blocks and cylindrical objects. The ideal grasp to pick up a block is to place each of the two fingers at the center of two opposing faces of the block, approaching perpendicularly to the surface. To minimize collision with other nearby blocks, the grasp is planned to approach from above. This greatly reduces the number of possible grasps and allows the user to skip a slower online planning phase, since we will already rely more on the user and less on an automated grasp quality analysis. Our system also allows us to manually design appropriate grasps for particular affordances that are difficult for an  automated  planner  to  recognize.  Our UI has been developed as a plugin for the GraspIt! \cite{miller2004graspit} simulator.

\subsection{sEMG Device Sensing and Control}

We have devised a control method which allows a user to control a cursor in 2 modes using a single sEMG sensor.  The sensor is shown attached to a subject in Fig. \ref{fig:semg_device_worn}, and its individual components are labeled in Fig. \ref{fig:device_semg}. The sEMG interface shown in Fig. \ref{fig:device_interface} maps a smoothed RMS value of a single EMG signal to one of several levels, where each level maps to a specific cursor action in the main UI as shown in Fig. \ref{fig:BCI_selection}.  A light muscle contraction (medium-level signal) causes the cursor to move between a set of options displayed on the screen.  A stronger muscle contraction (high-level signal) selects the highlighted option. This results in a simple yet effective interface that lets the user navigate and select actions from a menu. 
Further, the device has been refined to make the user feel in control of the system at all times by seeking to minimize the number of accidental responses recorded by way of spurious spikes in signal strength.  This is done by preventing the user from selecting a button unless they flex strongly.  Medium flexes are much more likely to happen accidentally than strong flexes, but they never cause any action to occur since they only change the currently highlighted button. 

\begin{figure}
	\centering
	\includegraphics[width=0.41\textwidth]{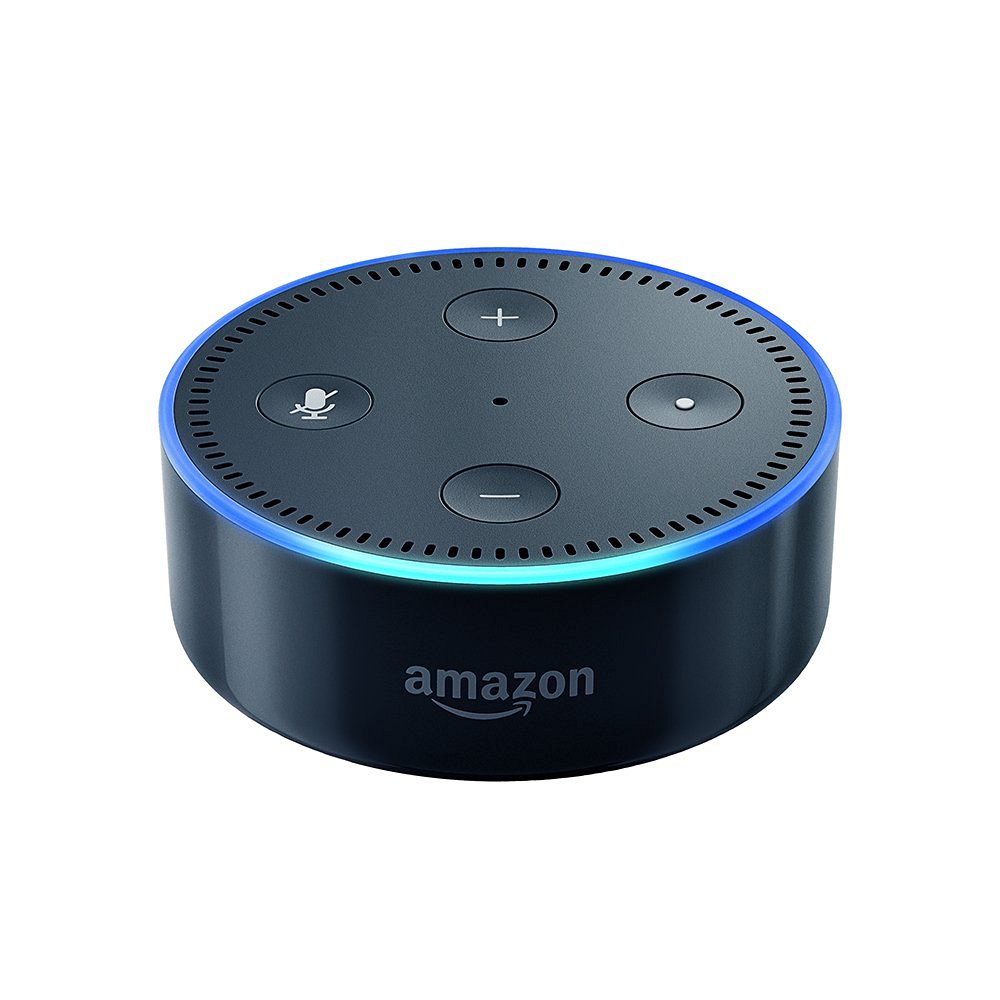}
	\caption{The Amazon Echo Dot allows a user to execute a command by speaking a trigger word followed by the associated program the user wants to run. For the Graspit! interface, users were asked to say ``Echo, tell the robot X'' where `X' was a command on the screen and `tell the robot' was the associated program.}\label{fig:device_echo}	
\end{figure}

\subsection{Speech Recognition Device}

We used a low cost Amazon Echo Dot and its Alexa API to implement a simple voice command activated user interface. This allows the user to say a phrase to execute a command that is available on the screen. For example, a user can say "Echo, tell the robot Rerun Vision", which would then execute that phrase - the button labeled 'Rerun Vision' will be selected. 

\begin{figure}[ht]
	\vspace{0.1in}
	\centering
	\includegraphics[width=0.3\textwidth]{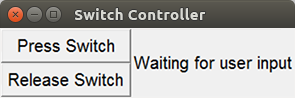}
	\caption{The Switch interface features clickable buttons to simulate the actions of the switch and displays the current action to be executed. In this case the system is waiting for user input.}\label{fig:switch_interface}
\end{figure}

\begin{figure}
	\centering
	\includegraphics[width=0.41\textwidth]{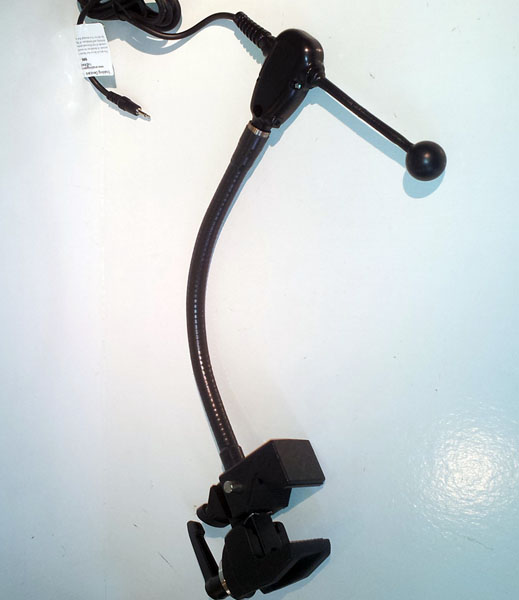}
	\caption{The Ultimate Switch featured a 10g of force toggle switch that could register a press from a release. The switch interfaced with a computer through a 3.5mm audio input port.}\label{fig:device_switch}
\end{figure}

\subsection{Ultimate Switch}

The Ultimate Switch is an adaptive switch that can activate with only 10 grams of force and is widely used by impaired individuals. We implemented an interface as shown in Fig. \ref{fig:switch_interface} that allows a user to navigate the menu with 2 modes of input based on timing. The switch can be pressed and released in any direction. A press followed by a release is registered as an input. The switch interface initially displays the message `Waiting for user input.' Once the switch is pressed, this message changes to `Going to send NEXT.' If the switch is released within a 0.1 - 1 second window, the input is registered as `next', highlighting the next button in the main interface. Pressing and holding for more than 1 second causes the message to become `Going to send Select.' Releasing the switch in a 1 - 3 second window selects the currently highlighted button. If the switch is pressed and released too quickly, or pressed and held for more than 3 seconds, the switch interface returns to the `Waiting for user input' state. Releasing the switch in this state will have no effect, and the user can restart the input.

\begin{figure*}[t]
	\vspace{0.1in}
	\centering
	\includegraphics[width=0.75\textwidth]{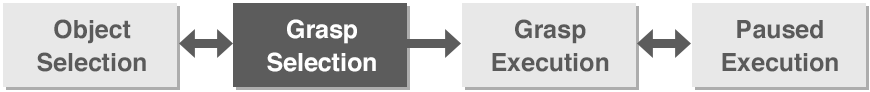}
	\caption{Execution Pipeline: The pipeline is placed at the top of each stage to clarify what stage the user is currently in, and what stage they will be sent to if they navigate forwards or backwards. In the Paused Execution state, the user is also given the ability to restart the entire system.} 
	\label{fig:pipeline} 
\end{figure*}

\begin{figure*}[t]
	\vspace{0.1in}
	\centering
	\begin{subfigure}[t]{.24\textwidth}
		\centering
		\includegraphics[width=0.98\textwidth]{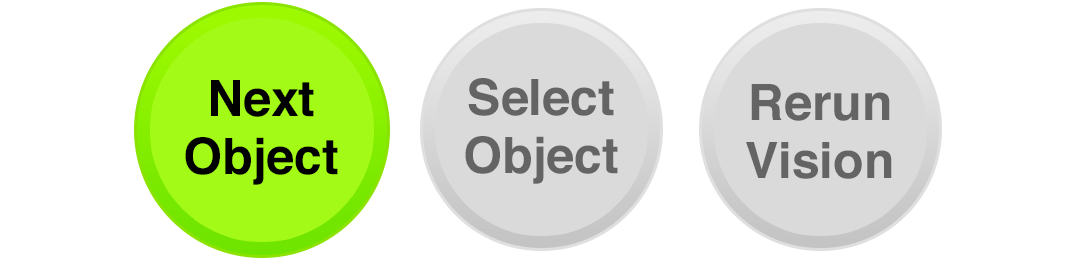}
		\caption{Object Selection}\label{fig:obj_sel}	
	\end{subfigure}
	\begin{subfigure}[t]{.24\textwidth}
		\centering
		\includegraphics[width=0.98\textwidth]{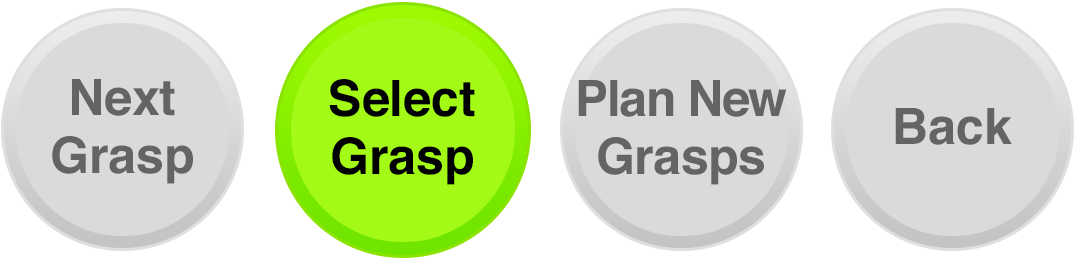}
		\caption{Grasp Selection}\label{fig:grasp_sel}
	\end{subfigure}
	\begin{subfigure}[t]{.24\textwidth}
		\centering
		\includegraphics[width=0.98\textwidth]{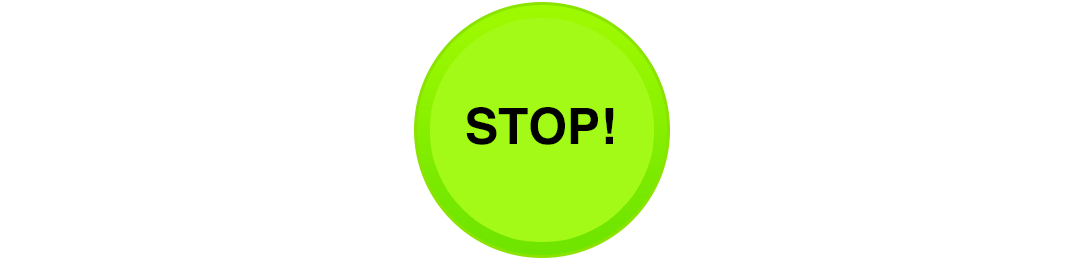}
		\caption{Grasp Execution}\label{fig:grasp_exec}
	\end{subfigure}
	\begin{subfigure}[t]{.24\textwidth}
		\centering
		\includegraphics[width=0.98\textwidth]{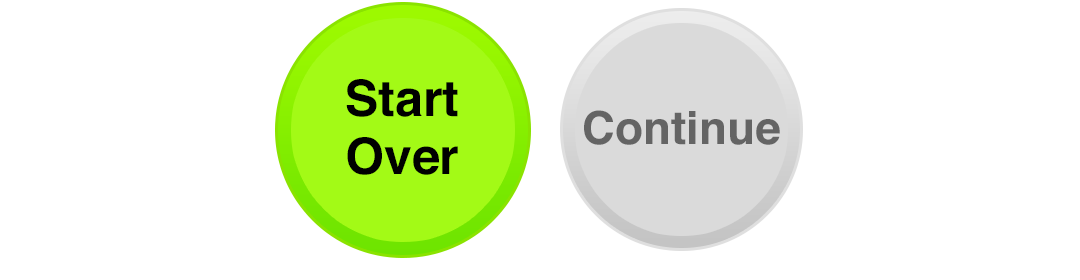}
		\caption{Paused Execution}\label{fig:paused}
	\end{subfigure}
	
	\caption{Stages to the Grasping Pipeline. Each stage has different options allowing the user to navigate the interface. While the user is navigating the system, the current option is highlighted green. A medium flex will change which button is highlighted, and a hard flex will select the currently highlighted button.} 
	\label{fig:grasping_pipeline} 
\end{figure*}

\subsection{Phases of the Planner}
\begin{figure}[t]
	\centering
	\begin{subfigure}{.1\textwidth}
		\includegraphics[width=1\textwidth]{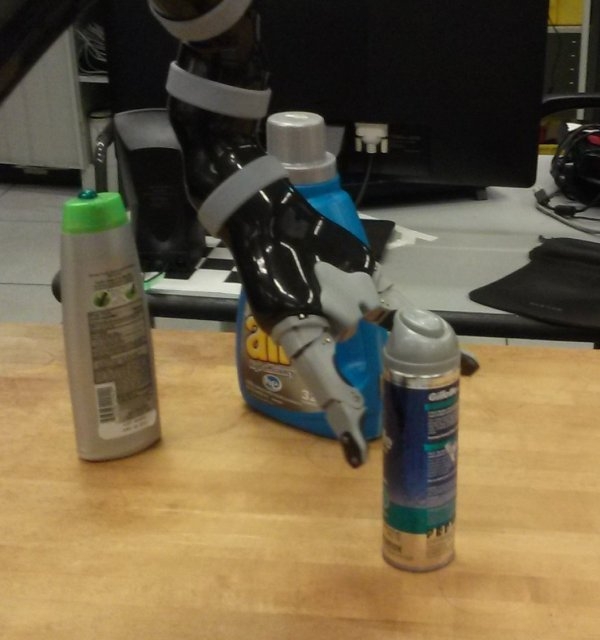}
		\caption{Approach}\label{fig:approach}	
	\end{subfigure}
	\begin{subfigure}{.1\textwidth}
		\includegraphics[width=1\textwidth]{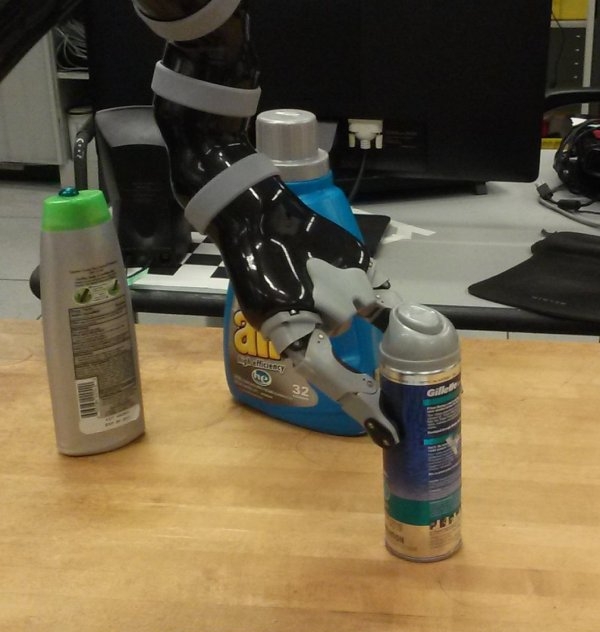}
		\caption{Grasp}\label{fig:grasp}
	\end{subfigure}
	\begin{subfigure}{.1\textwidth}
		\includegraphics[width=1\textwidth]{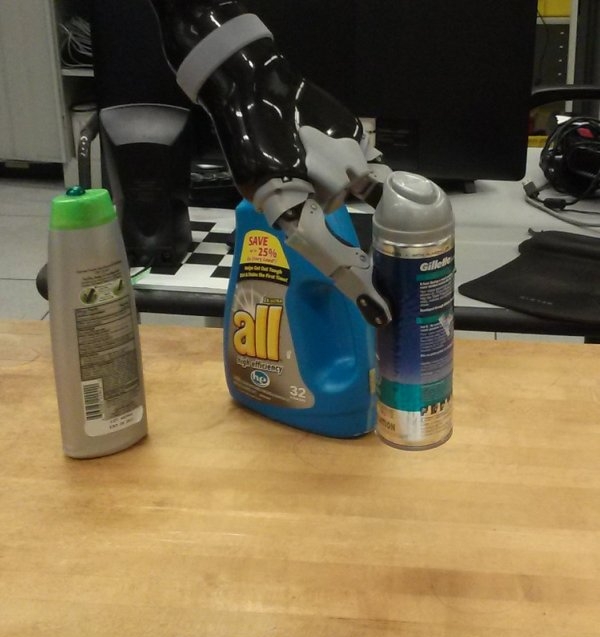}
		\caption{Lift}\label{fig:lift}	
	\end{subfigure}
	\begin{subfigure}{.1\textwidth}
		\includegraphics[width=1\textwidth]{img/grasp/grasp.jpg}
		\caption{Place}\label{fig:place}	
	\end{subfigure}
	\caption{Grasp Execution consists of 4 stages. First, the gripper is moved to a position that is slightly backed off from the final grasp position. The gripper then moves in towards the object and the fingers close. The object is then lifted off the table. Finally, the object is released in the target placement area after a horizontal translation.} 
	\label{fig:approach_grasp_lift} 
\end{figure}

The UI contains 5 states which the user navigates through in order to plan and execute a grasp on a target object. The buttons available in each state are shown in Fig. \ref{fig:grasping_pipeline}. The states consist of the following:
\\\noindent\textbf{Object Recognition: }
This stage initiates the object recognition system which detects objects in the scene using the point cloud and RGB image captured by a Microsoft Kinect. The recognition algorithm uses a RANSAC based approach to match point cloud data with mesh models of objects stored in a database\cite{papazov2010efficient}. While the recognition service is running in the background, the user sees the interface overlaid with an image indicating the status of the operation. The image fades away once the request is complete.  The system can identify multiple objects along with their location and orientation, and the next stage allows the user to identify their desired target object.
\\\textbf{Object Selection: }
This stage presents the scene populated with objects recognized from the object recognition stage. Here, user intent is dictated by three buttons dedicated to selecting the currently highlighted object, highlighting the next object in the scene, and rerunning the vision system to account for any new changes in the scene. The object currently selected is highlighted in green and all other objects that can be chosen next are gray. An example of what the UI looks like to the user in this stage is shown in Fig. \ref{fig:UI_views_obj}.
\\\textbf{Grasp Selection: } This stage presents the user with a set of possible grasps for the target object, displayed on the right side of the interface. The user can choose to either select the currently highlighted grasp or cycle through the available grasps until a preferable one is found. The user can also navigate back to the previous state in case the chosen object needs to be changed. In the background, the grasps are sent to MoveIt!\cite{chitta2012moveit} and are analyzed for reachability.  If the grasp is unreachable, then it is marked as red. If it is reachable, it is marked as green. Selecting a reachable grasp automatically confirms selection.
\\\textbf{Grasp Execution: } When this stage is entered, the selected grasp and its associated trajectories are sent to the arm to be executed using the process shown in Fig. \ref{fig:approach_grasp_lift}. While the chosen grasp is being executed, a single button is available for the user to select in order to pause the current execution procedure. This state is intentionally made very simple with a single button, so that the user is able to quickly stop the arm if required. 
\\\textbf{Paused Execution: } If the execution is stopped by the user, the system enters into the paused execution stage. Here, the user is presented with options to restart the system and go back to the Object Selection state or to continue with the currently paused grasp execution.

\subsection{Interface Improvements}

\begin{figure*}[t]
	\vspace{0.1in}
	\centering
	\begin{subfigure}[t]{.47\textwidth}
		\centering
		\includegraphics[width=1\textwidth]{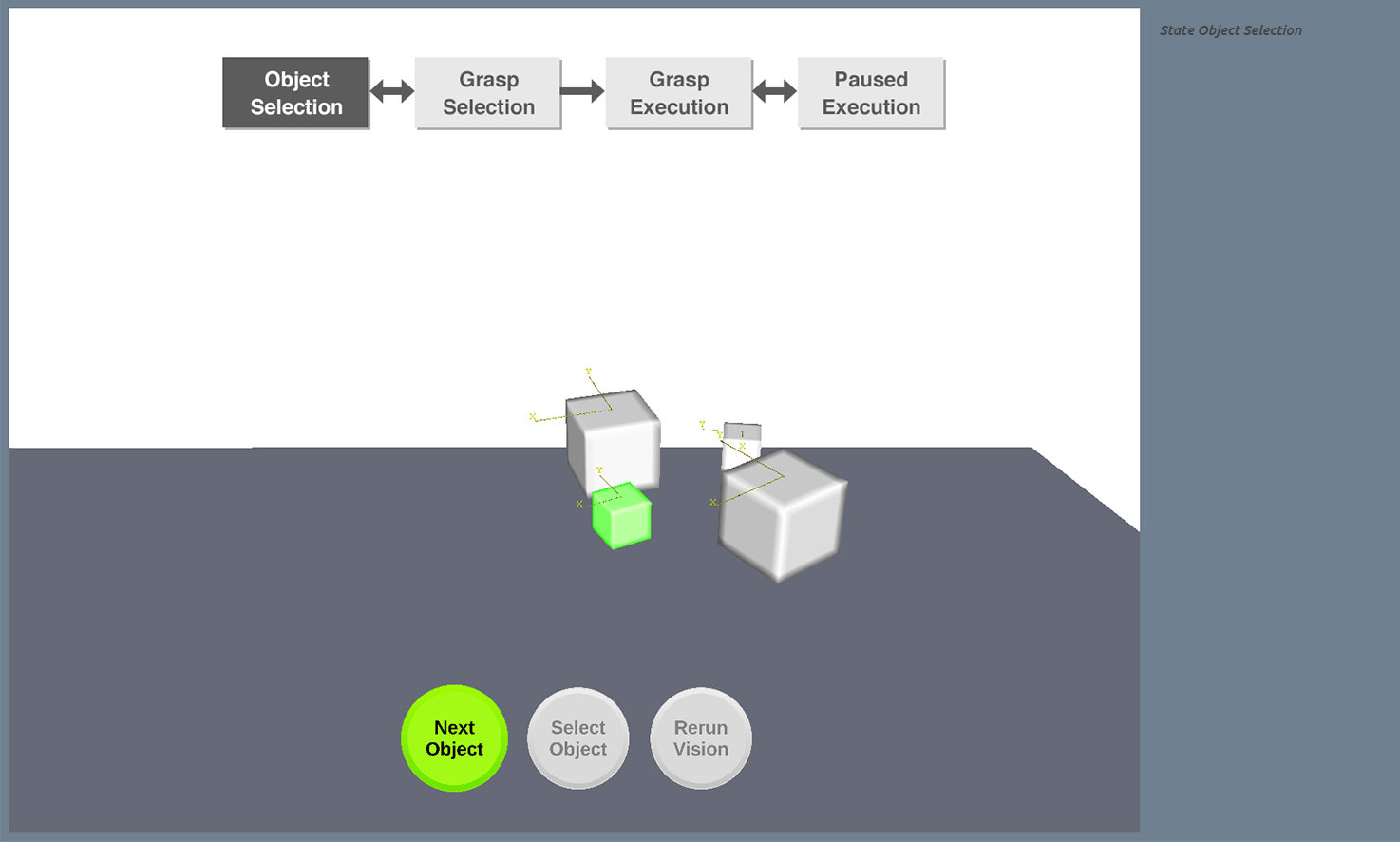}
		\caption{Object Selection Stage: Here, the currently selected object is highlighted in green, while the other detected objects are gray.}
		\label{fig:UI_views_obj}
	\end{subfigure}
	\hspace{.25in}
	\begin{subfigure}[t]{.47\textwidth}
		\centering
		\includegraphics[width=1\textwidth]{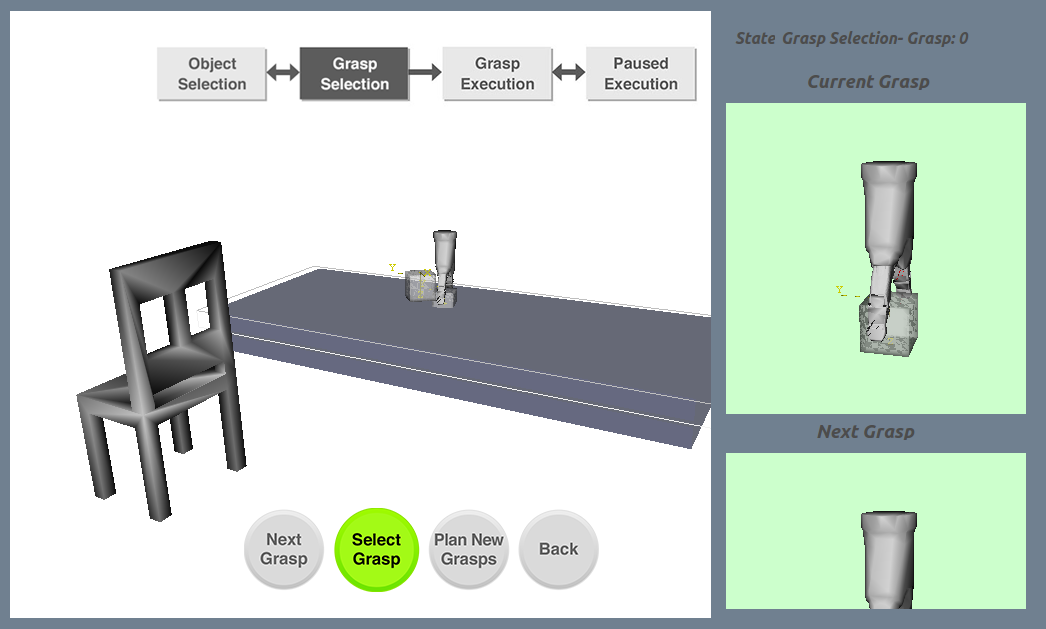}
		\caption{Grasp Selection Stage: The pipeline figure at the top shows which stage the user is currently at. The middle of the screen contains the planning scene. The current available options are displayed along the bottom. The images on the right show reachable grasps for the currently selected object highlighted in green.}
		\label{fig:UI_views_grasp}
	\end{subfigure}
	\caption{Stereotypical views of the user interface as displayed to the user.} 
	\label{fig:UI_views} 
\end{figure*}

We have improved the usability of the interface in several different ways.  
\\\textbf{Removed Point Cloud from Planning Scene: }Prior versions of the user interface had the point cloud of the scene overlaid on the planning scene. We found that while this was useful for developers to debug the system, it confused users as most were unfamiliar with point clouds and found it easier to view the scene directly in front of them.  In that line of thought, we have optimized this interface to work for someone who is able to view the scene directly in front of them rather than providing all information to the user through the interface.
\\\textbf{Oriented Planning Scene To User's Perspective: }Prior versions of the user interface showed the scene from the perspective of the Kinect which was capturing the scene. We rotated the planning scene within the user interface to show the scene from the same perspective the user is viewing it from, and added a chair to better indicate where the user is situated in space.  
\\\textbf{Pipeline Stage Diagram: }In order to help clarify how the user interface can be used, the pipeline stage diagram shown in Fig. \ref{fig:pipeline} is displayed above the planning scene. The pipeline stage diagram shows what stage the user is currently in, what will come next and what they have already completed. This makes the consequences of different button selections significantly more clear. For example, it is now clear to a user that they are in the grasp selection stage, and that the next stage should they click to move forward is the execution stage which will actually cause the arm to move. 
\\\textbf{Running Recognition Notification: }A large notification is displayed when object recognition is running.  It is the only stage of the pipeline where the user is not able to provide input.  As this process takes several moments, it is helpful to make it clear to the user that they just need to wait until object recognition is finished.
\\\textbf{Improved Interface design: }The previous design of the interface was improved in order to be less distracting and confusing to the user. Only two colors are now used, instead of blue, green, and red. Green indicates an active selection of an object or button, and gray indicates an inactive object or button. The buttons and labels have also been enlarged and centered on the screen for better readability.
\\\textbf{Enhanced and Accessible Code Base: }Several modifications were made to the system to improve its reliability and also to simplify the development process. The most prominent of these changes was the decision to isolate the interface and the sEMG interaction component from the core GraspIt! Application code and repackaging it into a lightweight plugin. This improved the scalability of the system from the developer's perspective as it allowed changes to be made to the underlying system at a rapid pace.  The GraspIt! plugin code is available on Github along with detailed setup instructions. Several stereotypical views of the UI are shown in Fig. \ref{fig:UI_views} demonstrating all of the above changes.

\section{User Study}

In order to explore the performance of the new system, we designed an experiment around the use of the four input devices with our system, namely the mouse, Ultimate Switch, Amazon Echo Dot with Alexa speech recognition, and our novel sEMG device. All testing was approved by the Institutional Review Board of Columbia University under Protocol IRB-AAAJ6951.

15 human subjects were trained and then timed using the four input devices into the system. The subjects had never used the system prior to their participation in this study. The scene initially contained 3 target blocks (2 inch, 2.5 inch, and 3 inch blocks). Once those three target blocks were moved, the user was then tasked with moving a shaving cream can to demonstrate that the system can also handle arbitrary objects like those in the YCB object database\cite{YCBDataset}. Each subject ran 4 pick and place operations for each of the four input devices, totaling 16 pick and place operations.

The training process consists of several steps: (1) a system tutorial where the user navigates through the system using a computer mouse while the user interface is explained. The time taken to learn and navigate the system during this tutorial is measured. They are then (2) subsequently trained on the Amazon Echo and the Ultimate Switch. To prevent order effects, each subject is randomly assigned to use the switch first, followed by the Echo Dot, or vice versa. The user is then (3) trained on the sEMG device through a calibration step and they learn how to consistently produce either a medium or high signal at will. During this step, the gains and thresholds are tuned, and the electrodes may be repositioned to improve the signal. This normally takes 4-5 minutes for both the forearm and the ear. After the subject is trained on each device, we have the subject use the system 4 times, once for each target block and once for the shaving cream can. A video and a description of the protocol used for this study is available with the source code used at \url{http://crlab.cs.columbia.edu/HumanRobotInterfaceforAssistiveGrasping/}.

\begin{figure}
	\centering
	\includegraphics[width=.47\textwidth]{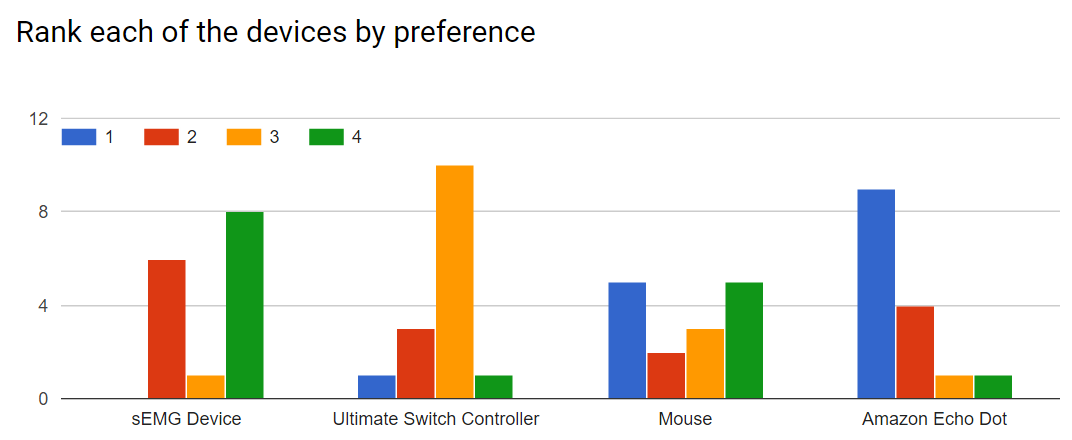}
	\caption{A qualitative preference of the users in our study with the devices they used in the experiment. In general users showed a strong preference for speech recognition and mouse input}
	\label{fig:my_label}
\end{figure}

\begin{table*}[t]
	\centering
	\begin{tabular}{|c|c|c|c|c|}
		\hline
		\multicolumn{5}{|c|}{Average time per input device} \\ \hline
		Activity & Mouse (s)	& Alexa (s)	& Switch (s) & sEMG (s) \\ \hline
		explain interface & 169.58 & 80.76 & 78.97 & 314.15 \\ \hline
		user block 1 & 20.5 & 18 & 20.5 & 18.0 \\ \hline
		robot block 1 & 63.7 & 40.87 & 48.05 & 53.01 \\ \hline
		user block 2 & 33.43 & 23 & 38.5 & 27.3 \\ \hline
		robot block 2 & 60.39 & 53.52 & 49.548 & 50.07 \\ \hline
		user block 3 & 21.5 & 17 & 28 & 13 \\ \hline
		robot block 3 & 61.18 & 60.39 & 74.335 & 68.27 \\ \hline
		user YCB bottle & 23 & 74.14 & 77.09 & 124.24 \\ \hline
		robot YCB object & 73.39 & 40.91 & 67.71 & 65.7 \\ \hline
	\end{tabular}
	\caption{ Average times for both the time taken for the user to select and object and grasp, followed by the time taken by the robot to grasp the object. The robot was consistent in taking roughly 50-70 seconds to execute a grasp.  \label{timingdata}}
	
\end{table*}

\begin{table*}[t]
	\centering
	\begin{tabular}{|c|c|c|c|c|c|c|}
		\hline
		\multicolumn{7}{|c|}{Number of successful grasps} \\ \hline
		Activity & Mouse 15 trials & Alexa 15 trials & Switch 15 trials & sEMG (forearm) 7 trials & sEMG (behind ear) 8 trials & Average \\ \hline
		Block 1 & 100\% & 100\% & 100\% & 100\% & 100\% & 100\% \\ \hline
		Block 2 & 100\% & 100\% & 100\% & 100\% & 100\% & 100\% \\ \hline
		Block 3 & 100\% & 100\% & 100\% & 100\% & 100\% & 100\% \\ \hline
		YCB Object & 66.67\% & 80\% & 80\% & 71.43\% & 87.50\% & 76.53\% \\ \hline
		Average & 92\% & 95\% & 95\% & 93\% & 97\% & 94.13\% \\ \hline
	\end{tabular}
	\caption{ Success rate of grasping objects during each trial. Each trial occurred 15 times and shows the number of successful grasps as well as the percentage successful.  }
	\label{tab:trialdata}
	
\end{table*}

\subsection{Lessons Learned}

For the user study, a trial was considered successful if the user grasped an object and carried it over to the other location on the table. A failure was when the arm could not recognize the object after three attempts or if the arm failed to pick and place the object. The subject used the mouse first in the experiment consistently and therefore the success rates were much lower than the other input devices, which we attribute to an unfamiliarity with the interface. We randomly assigned placement of the sEMG device to either behind the subject's ear or on their forearm. Users who had placed the sEMG device behind their ear were slightly more successful than those who had placed it on their forearm. An object form the YCB object database\cite{YCBDataset} was assigned to each user to grasp in the final stage of the trial for each input device. Users had no difficult picking up various sized cubes but showed reduced performance with the YCB object. On average users were successful at picking up an object 94.13\% of the time.

The timing results are shown in Table \ref{timingdata}. The major take away from this is that most users had very little difficulty using any of the four inputs devices. The times listed under YCB object trials took much longer than those for the blocks because the time taken to calculate the grasps was so much longer than it was for the blocks. This inflated the time spent waiting to choose a grasp for the given object. Occasionally users would have difficult with the interface such as recalibration of the sEMG device or a peripheral crashing. In cases such as these the timer was reset and the subject was asked to redo the experiment. 

In addition to this added grasp calculation time, users often took much more time with the YCB object because of trouble calculating any valid grasps due to point cloud error. In these circumstances we had the user rerun object recognition on the scene to attempt to find a better view of the object. Moving forward we will likely rerun vision automatically in the event that no valid grasps are found and improve our methodology for finding grasps. All four devices had similar times for all of the trials and we take this information to show that using an sEMG device is as fast or in some cases faster than using the other three input devices, and therefore validates it as a useful input device into a HitL system. 

One important thing to note is that the training time for the mouse and sEMG device were the longest. The mouse time included an introduction into the system, which inflated the amount of time it took to train the user in this case. The mouse was also the first device the user had a chance to use. The sEMG device took additional time to train the user since it had to be calibrated. Training for the sEMG device also involved finding a valid electrode location, showing users the interface, and walking them through how the device would control the interface. Training users on the Amazon Echo and the Ultimate Switch took substantially less time as they were easy to use and very similar in use to the mouse. However once the user became familiar with each device the timings were all comparable. 

Overall, the users understood how they were supposed to navigate the system and were aware of what the system was doing at any given point in time. From our user studies we were able to both verify that our improvements to the new system had the intended effect, and several new areas of improvement became apparent. 

One lesson learned from the experiments was that displaying more of the current scene would be useful to the user, for example showing an overlaid image of the scene on top of the user interface was a suggestion given by several of the participants. Another suggestion was that showing the perspective of the objects relative to the user would have been helpful. Many users were confused by the orientation of the grasp relative to the object and found the display to be relatively counter intuitive.

\subsection{Subject Survey}
After completion of the user study, the subjects filled out a short questionnaire about their experience with the system.  From the survey,  it was apparent that many users found the system easy to use and 86.7\% of participants reported that they found the task useful and would use it in their everyday life. 80\% reported that they would use the speech recognition device through the Amazon Echo in their everyday life. When considering future applications of these devices we took these results to indicate that able-bodied individuals find speech recognition to be the easiest system of the four to use for input into the robotic interface. 

\section{Conclusion and Future Work}
This work has presented a new and improved interface for assistive robotics. The interface provides a higher level of autonomy than before.  It is also easier for users to understand how to effectively interact with the system due to improvement in how both the state of the system and consequences of a user's actions are displayed. We have also demonstrated that the system can be used easily with a variety of input devices and the amount of time to execute an action is comparable across multiple input devices. This system is being migrated to the Columbia University Medical Center in order to support extensive user studies on subjects with spinal cord injuries with the goal of providing assistance with activities of daily living. 

\bibliographystyle{abbrv}
\bibliography{bib/bci_refs_2,bib/grasp_bib,bib/robotics}

\end{document}